\documentclass[pdftex]{article}
\usepackage{icrctc07}

\title{Observations of Gamma-ray Bursts with VERITAS and Whipple}
\shorttitle{GRB Observations at VERITAS \& Whipple}
\authors{D. Horan$^{1}$ for the VERITAS Collaboration$^{2}$.}
\shortauthors{D. Horan et. al}
\afiliations{$^1$Argonne National Laboratory, 9700 S. Cass Ave.,
Argonne, IL 60439, U.S.A.\\ $^2$ For full author list see G. Maier
``VERITAS: Status and Latest Results'', these proceedings}
\email{deirdreh@hep.anl.gov}

\abstract{Many authors have predicted very-high-energy (VHE; {\it{E}}
$>$ 100 GeV) emission from gamma-ray bursts (GRBs) both during the
prompt phase and during the multi-component afterglow. To date,
however, there has been no definitive detection of such
emission. Recently, the Swift Satellite made the exciting discovery
that almost 50\% of GRBs are accompanied by one or more X-ray flares,
which are found to occur from several seconds to many hours after the
prompt emission. The discovery of this phenomenon and the many
predictions that VHE emission should accompany these flares increases
the already strong motivation for making immediate follow-up VHE
observations of GRBs. Observations of GRBs have high priority at
VERITAS, preempting any observations that may be in progress. GRB
alerts are received from the GCN via a socket connection. This is
interfaced to the VERITAS Tracking Software to minimize the time
between a notification arriving and the telescope being slewed to the
GRB. We report here on GRB observations with VERITAS and with the
Whipple Telescope from 2005 through 2007.}

\begin{document}
\maketitle

\section{Introduction}

Gamma-ray bursts (GRBs) have been well studied at all wavelengths
since their discovery in 1969 \cite{Klebesadel:73}. The
very-high-energy (VHE; {\it{E}} $>$ 100\,GeV) band is the only energy
regime in which definitive evidence for GRB emission has yet to be
detected. For the observation of photons of energies above 100\,GeV,
only ground-based telescopes are available at present. These fall into
two broad categories, air-shower arrays and atmospheric Cherenkov
telescopes (of which the majority are Imaging Atmospheric Cherenkov
Telescopes). The air shower arrays, which have wide fields of view
making them particularly suitable for GRB searches, are relatively
insensitive. There are several reports from these instruments of
possible TeV emission \cite{Padilla:98:Airobicc} \&
\cite{Amenomori:01:TibetGRBs}. The Milagro Collaboration reported on
the detection of an excess gamma-ray signal during the prompt phase of
GRB\,970417a with the Milagrito detector \cite{Atkins00}. In all of
these cases, the statistical significance of the detection is not high
enough to be conclusive.

The Milagro Collaboration have performed a number of searches for VHE
emission from GRBs \cite{Abdo:06:ShortGRBs},
\cite{Atkins:05:MilagroGRBCounterparts} \&
\cite{Atkins:04:MilagroGRB}, but no evidence for VHE emission was
found from any of these searches. Atmospheric Cherenkov telescopes,
particularly those that utilize the imaging technique, are inherently
more flux-sensitive than air-shower arrays and have better energy
resolution but are limited by their small fields of view
(3\,-\,5$^\circ$) and low duty cycle ($\sim$\,7\%). In the era of the
Burst And Transient Source Explorer (BATSE), attempts at GRB
monitoring were limited by slew times and uncertainty in the GRB
source position \cite{Connaughton:97}. Recently, upper limits on the
VHE emission from the locations of seven GRBs observed with the
Whipple 10m Telescope were reported \cite{Horan:2007}. With their
fast-slewing telescope and low energy threshold, the MAGIC Group have
placed the most stringent upper limits on VHE emission from GRBs to
date \cite{Albert:06b} \& \cite{Albert:06a}.

\begin{table*}
\begin{center}
\begin{tabular}{l|ccccccc}
\hline
             & Discovery  & GCN          &  \multicolumn{2}{c}{{Best Location (J2000)}}  & T$_{90}$$^{a}$ & Fluence$^b$ &     \\
GRB          & Instrument & Number       & R.A.            & Dec.                        & (sec)            & (x10$^{-6}$)& $z$ \\
\hline
060501       & BAT         & 5040        & 21h 53m 30s  & +43$^\circ$ 59' 53''           & 26               & 1.2$\pm$0.1 & N/A\\
061222a      & BAT         & 5954        & 23h 53m 03s  & +46$^\circ$ 31' 59''           & 72               & 8.3$\pm$0.2 & $<$3\\
070311       & IBAS        & 6189        & 05h 50m 08s  & +03$^\circ$ 22' 30''           & 50               & 2           & low-$z$$^c$\\
070419a      & BAT         & 6302        & 12h 10m 59s  & +39$^\circ$ 55' 34''           & 116              & 0.6$\pm$0.1 & 0.97\\
070521       & BAT         & 6431        & 16h 10m 39s  & +30$^\circ$ 15' 22''           & 38               & 8.0$\pm$0.2 & 0.553$^d$\\
070612b      & BAT         & 6511        & 17h 26m 54s  & -08$^\circ$ 45' 06''           & 14               & 1.7$\pm$0.1 & N/A \\
\hline
\end{tabular}
\caption{\label{grb-positions}The gamma-ray bursts.}
\end{center}

$^a$T$_{90}$ is the time during which 90\% of the prompt emission was
detected.  The BAT T$_{90}$ is measured between 15-350 keV while the
IBAS T$_{90}$ is measured between 20-200 keV. $^b$ The BAT fluence is
measured between 15-150 keV while the IBAS fluence is measured between
20-200 keV.  $^c$ \cite{GCN:6203} propose that this GRB is at low
redshift (see text).  $^d$ A galaxy at $z$=0.0307 was also found in
the XRT error circle but it is less likely to be associated with the
GRB.

\end{table*}

\section{The Gamma-Ray Bursts}

Six different GRB locations were observed with VERITAS between March
2006 (when VERITAS was first operated as a 2-Telescope array) and June
2007. The properties of these bursts are summarized in
Table~\ref{grb-positions}. For three of these GRBs, all of the data
were taken at large zenith angles (elevation $<$ 50$^{\circ}$); the
detailed analysis of these data will be presented at a later
date. Upper limits on the VHE emission from the three bursts observed
at high elevation will be presented here. Those GRBs are described in
the following subsections.

Eleven GRB locations were observed with the Whipple 10m Telescope
between January 2005 and May 2007. The detailed description of these
bursts and their analysis will be presented in \cite{Dowdall:2007}.

\subsection{GRB 070311}

This GRB was detected at 11:52:50 UT by the IBAS in the IBIS/ISGRI
data \cite{GCN:6189}. It had a peak flux of 0.9 photon cm$^{-2}$
s$^{-1}$. The Swift XRT light curve followed a single power law with
an index of 1.9 $\pm$ 0.4 from 7 to 14 ks after the burst
\cite{GCN:6192}. The robotic 20-cm REM telescope at La Silla detected
optical emission from the position of the GRB with an approximate
magnitude of 14.3 about 3 minutes after the burst
\cite{GCN:6190}. This detection was confirmed by the 1.3m PAIRITEL
project \cite{GCN:6191} and at a number of other telescopes
\cite{GCN:6196}, \cite{GCN:6195} \& \cite{GCN:6198}. When the optical
afterglow was observed with the MDM 1.3m telescope on 070313, it was
found to have brightened by 0.99 mag since the previous night
\cite{GCN:6203}. The authors suggested that, given the level of
Galactic extinction at this GRB location, it is possible that this
burst was at low redshift and that the brightening was the result of
the onset of a supernova. Further monitoring of the afterglow
\cite{GCN:6208} found that the re-brightening peaked on day 2 after
which it declined rapidly \cite{GCN:6219}. It was proposed that the
optical re-brightening of this burst was due to late central engine
activity \cite{GCN:6209}. The GRB was not detected with the VLA
\cite{GCN:6207}.

\subsection{GRB 070521}

This burst was detected at 06:51:10 UT by the BAT on Swift (trigger
279935) \cite{GCN:6431}. The time-averaged spectrum from T-14.5 to
T+49.7 seconds was best fit by a power law with an exponential cutoff
and photon index of 1.10 $\pm$ 0.17 and peak energy of 195 $\pm$ 123
keV. The 1-sec peak photon flux measured from T+30.48 seconds in the
15 - 150 keV band was 6.7 $\pm$ 0.3 photon cm$^{-2}$ s$^{-1}$
\cite{GCN:6440}. The XRT light curve exhibits initial flaring
behaviour superposed on the power-law decay up to $\sim$T + 600s
\cite{GCN:6452}. \cite{GCN:6433} reported that the XRT error circle
coincided with the outskirts of a nearby ($z$=0.0307)
galaxy. Observations with the Subaru Telescope \cite{GCN:6444}
revealed a faint source inside the XRT error circle at a redshift of
$z$=0.553, and this source was also detected weakly with the Gemini
North \cite{GCN:6450} and Keck Telescopes \cite{GCN:6451}. Two other
sources consistent with the XRT error circle were also detected with
Keck. Observations with Gemini North on 070522 detected marginal
evidence for fading of the $z$=0.553 source \cite{GCN:6457} as well as
a detection of the third source detected by the Keck. The strength of
this third source was fainter than that reported by the Keck, but the
authors caution that the Gemini photometry at this location is quite
uncertain due to the presence of a nearby bright galaxy.

\subsection{GRB 070612b}

This burst was detected at 06:21:16 UT by the BAT on Swift (trigger
282073) \cite{GCN:6511}. The time-averaged spectrum from T-6.4 to
T+10.3 seconds was best fit by a simple power law with a photon index
of 1.55 $\pm$ 0.11. The 1-sec peak photon flux measured from T+10.84
seconds in the 15 - 150 keV band was 2.6 $\pm$ 0.4 photon cm$^{-2}$
s$^{-1}$ \cite{GCN:6523}. Although the XRT did detect an x-ray source,
no evidence for x-ray flaring was seen \cite{GCN:6521}. Many optical
telescopes performed follow-up observations of this GRB but no optical
emission was detected \cite{GRB070612b}.

\section{The VHE GRB Data and Analysis}

Burst notifications are received over a socket connection at both
VERITAS and the Whipple 10m Telescope. GRB observations take priority
over all other observations at both instruments so, whenever an
observable burst is detected, the telescopes are immediately
re-pointed and data are taken on the GRB location until the burst is
more than three hours old. VERITAS and the Whipple Telescope can slew
at 1$^\circ$s$^{-1}$ thus reaching any part of the observable sky
within 6 minutes.

\begin{table*}
\begin{center}
\begin{tabular}{l|ccccc}
\hline
              & $\Delta$T$^a$ & Exp.$^b$  & No. of     & 99\% Flux Upper & Time span of     \\
GRB           & (min.)        & (min.)    & Telescopes & Limit (Crabs)   & U. L.$^c$ (min.) \\
\hline
070311         & 44           & 110       & 2          & 3.7\%            & 44 - 124 \\
               & 5805         & 40        & 3          & 6.4\%            & 96.8 - 97.4$^d$ \\
070521         & 18           & 80        & 3          & 2.9\%            & 18 - 98 \\
070612b        & 24           & 80        & 3          & 2.0\%            & 24 - 151 \\
\hline
\end{tabular}
\caption{\label{grbs_VERITAS}The VERITAS GRB Observations.}
\end{center}

$^a$ The time in minutes between the start of the GRB and the
beginning of observations. $^b$ The total VERITAS exposure time on the
GRB. This is not necessarily one contiguous block. $^c$ For the
analysis presented here, all data for each GRB were combined to give
one upper limit. This column gives the duration after T$_0$ for which
the upper limit pertains. $^d$ The time is quoted in hours for this
observation.

\end{table*}

The VERITAS array is described in detail in
\cite{Maier:ICRC2007-VERITAS}. The GRB data were taken in both
{\it{wobble}} and {\it{tracking}} modes \cite{Daniel:ICRC2007}. At all
times during data-taking, the point-spread function of VERITAS was
$<$0.1$^\circ$ (the field of view of one PMT is 0.15$^\circ$). For the
three GRBs reported upon here, the positional offset between the final
GRB location and the position tracked by VERITAS were such that a
conventional ``point source'' analysis could be performed.

The data have been analysed using independent analysis packages (see
\cite{Cogan:ICRC2007}, \cite{Daniel:ICRC2007} for details on the
analyses). All of these analyses yield consistent results. The
calibration techniques are described in \cite{Hanna:ICRC2007}. The
99\% confidence level VERITAS upper limits for these GRBs are given in
Table~\ref{grbs_VERITAS}.

\section{Discussion and Conclusions}

\begin{figure}
\begin{center}
\includegraphics [width=0.35\textwidth]{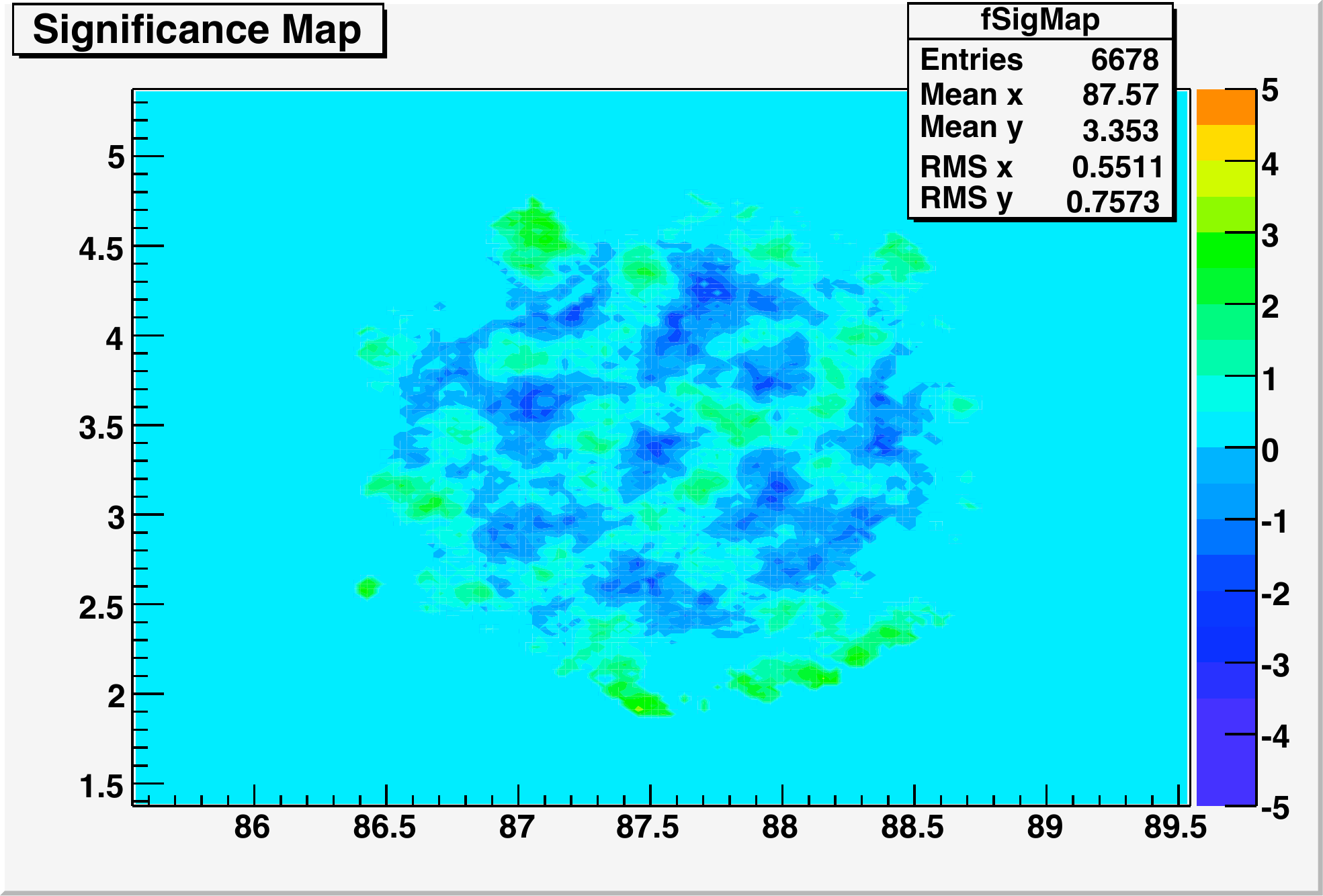}
\includegraphics [width=0.35\textwidth]{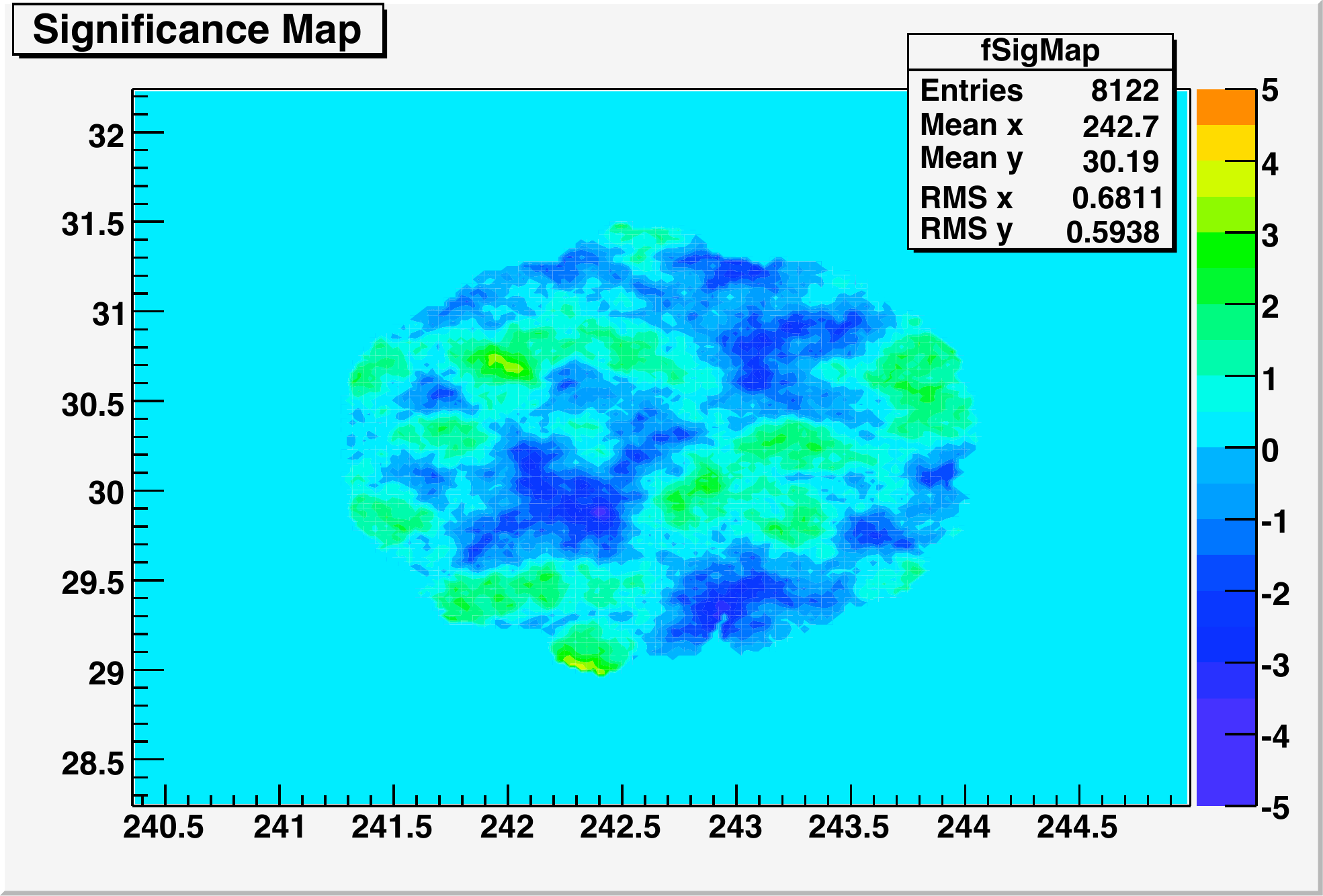}
\includegraphics [width=0.35\textwidth]{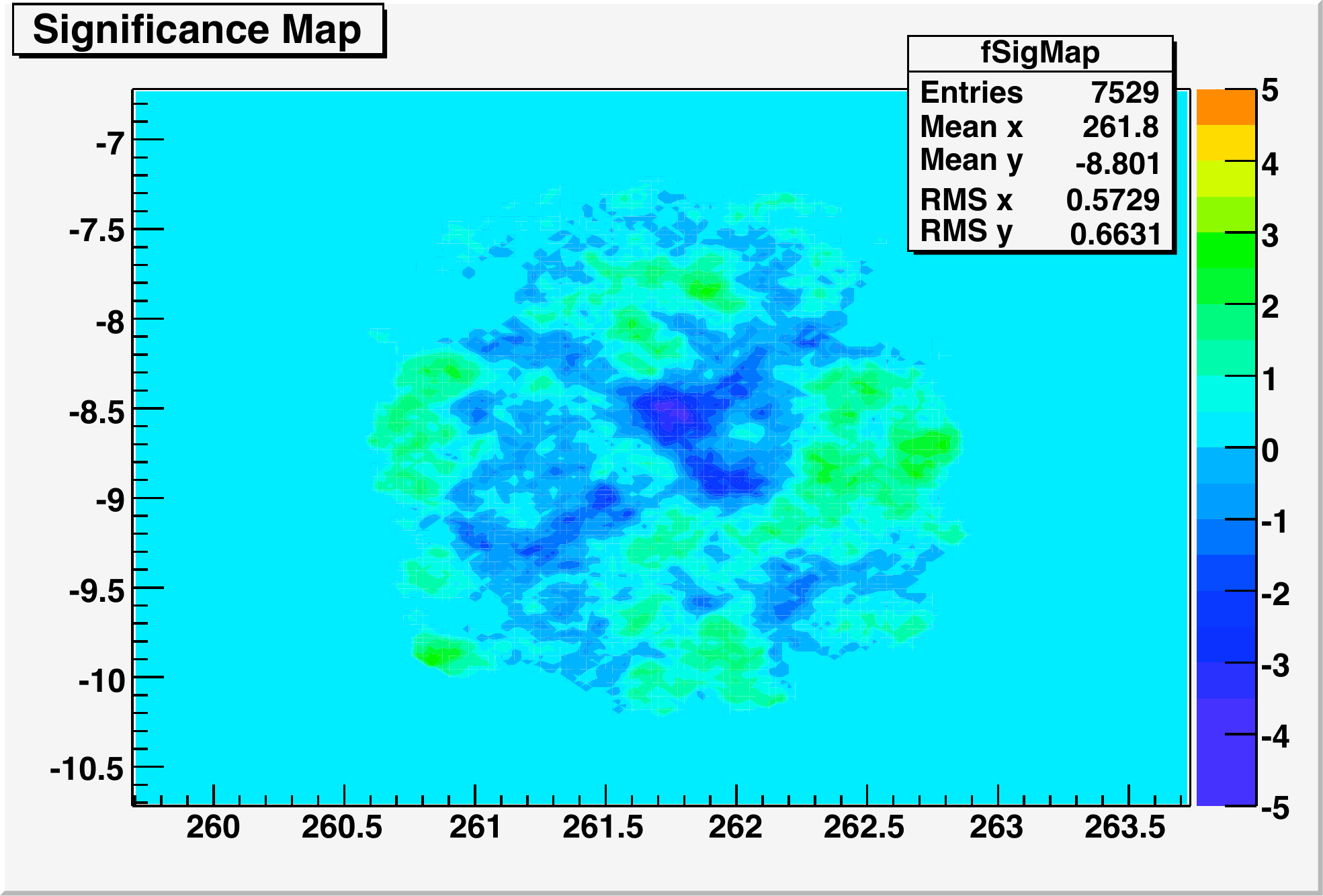}
\end{center}
\caption{\label{2dmaps}Two-dimensional significance maps for the three
GRB locations presented here; top: GRB 070311 (on UT 070311); middle:
GRB 070521; bottom: GRB 070612b.}
\end{figure}

Upper limits (99\% confidence level) for three of the six GRBs
observed to date with the VERITAS array were presented here. The
limits range from 2.0\% to 6.4\% of the Crab Nebula flux at a peak
response energy of approximately 300 GeV. These data were taken in 2-
and 3-Telescope observing mode during 2007. Analysis of the remaining
three GRBs is ongoing. So far, the upper limits presented span the
entire duration of the gamma-ray burst
observation. Figure~\ref{2dmaps} shows the VERITAS two-dimensional
significance maps for the three GRB locations. Only one of the GRBs
observed here, GRB 070521, has a distance measurement, which places it
at a redshift of 0.553. It is likely that GRB 070311 occurred at low
redshift but no definitive distance measurement was obtained. There
was no optical afterglow detected from GRB 070612b so its redshift is
unknown. None of the upper limits presented here have been corrected
for the effects of absorption on the extragalactic background light.

A detailed report of the Whipple 10m GRB observations and their
analysis will be presented in \cite{Dowdall:2007}. No evidence for TeV
emission was found during these Whipple observations.

\section{Acknowledgments}

This research is supported by grants from the U.S. Department of
Energy, the U.S. National Science Foundation, and the Smithsonian
Institution, by NSERC in Canada, by PPARC in the UK and by Science
Foundation Ireland.

\nocite{GCN:5984}
\bibliography{icrc0406}
\bibliographystyle{plain}

\end{document}